\documentstyle[aps]{revtex}
\input epsf.sty
\begin{document}
\twocolumn[\hsize\textwidth\columnwidth\hsize\csname@twocolumnfalse\endcsname
\title{The Van der Waals interaction of the hydrogen molecule - \\
an exact local energy density functional.}
\author{T.C. Choy}
\address{
National Center for Theoretical Sciences (Physics Division), \\
P.O. Box 2-131, Hsinchu, Taiwan 300, R.O.C. and \\ School of
Physics, University of Melbourne,\\ Parksville, Melbourne,
Victoria 3001, Australia}
\date{\today}
\maketitle
\begin{abstract}
We verify that the van der Waals interaction and hence {\it all}\
dispersion interactions for the hydrogen molecule given by:
\begin{equation}
W^{\prime\prime}= -{A \over R^6}-{B \over R^8}-{C \over R^{10}}-
\dots, \label{Heqn1a}
\end{equation}
in which $R$ is the internuclear separation, are exactly soluble.
The constants $A=6.4990267 \dots$, $B=124.3990835 \dots$ and
$C=1135.2140398 \dots$ (in Hartree units) first obtained
approximately by Pauling and Beach (PB) \cite{PaulingBeach35}
using a linear variational method, can be shown to be obtainable
{\it to any desired} accuracy via our exact solution. In addition
we shall show that a {\it local} energy density functional can be
obtained, whose variational solution rederives the exact solution
for this problem. This demonstrates explicitly that a {\it static}
local density functional theory exists for this system.  We
conclude with remarks about generalising the method to other
hydrogenic systems and also to helium.
\end{abstract}
]  \narrowtext

\section{Introduction}
\label{Intro} Amongst the few non-trivial many-body problems in
quantum mechanics, the hydrogen molecule was the first system to
be thoroughly studied and continues to be researched, motivated by
experiments \cite{Boudart72},\cite{Chin96},\cite{Kleppner99} and
by advances in modern LDA techniques of computation
\cite{Perdew96}. It is perhaps not so well known that the
dispersion forces for this elementary system is amenable to an
exact solution, due to certain confusion in the early literature
with regard to the methods used to attack it.  In this paper we
shall use a method first propounded by Slater and Kirkwood (SK)
\cite{SlaterKirkwood31}, but whose equation (see eqn(\ref{Heqn8})
below) they were unable to solve, to show that the dispersion
forces for the hydrogen molecule are exactly soluble. This method
later became formalised as the method of Dalgarno and Lewis
\cite{DalgarnoLewis55} and is particularly suited for the problem
of dispersion forces in general. However it does not seem to have
been exploited in recent studies of the van der Waals interaction
using density functional theories, \cite{DobsonDW98},
\cite{KohnMM98} . We shall demonstrate the superior convergence
properties of the SK method. This was already known to the early
pioneers, \cite{PaulingBeach35}, \cite{SlaterKirkwood31}, in
contrast to frequency dependent methods, essentially based on a
summation over dipole matrix elements with excited states or an
integration over the dynamical polarizabilities. The latter was
derived through the original work of Eisenchitz and London
\cite{EisenchitzLondon30}, and had a rather strong influence on
later studies \cite{MahantyNinham76}. For the hydrogen system, we
shall derive an {\it exact} local density functional theory,
soluble for this case, and whose solution converges to the exact
results. We shall show that the method is generalisable to other
hydrogenic systems and also to helium for which systematic
approximations can be derived. In section \ref{Background}, we
shall discuss the background to the exact solution verifying that
the PB variational method is essentially exact, though more slowly
convergent than the SK method. In section \ref{Exact} we shall
present our method for solving the SK equation and show our
results. In section \ref{DFT} we shall formulate the local density
functional theory (LDFT), discuss its solution and show that it
converges to the exact results of section \ref{Exact}. In section
\ref{Helium} we shall discuss the problem of helium and conclude
in section \ref{Conclusion} with discussions about further work.
\section{Background}
\label{Background} The Hamiltonian for the dispersion forces of
the hydrogen molecule was first derived by Margenau
\cite{Margenau31} and has appeared in subsequent editions of many
textbooks, especially the famous one of Pauling and Wilson
\cite{PaulingWilson35}.  The latter contains an excellent survey
of the variational treatment for the van der Waals interaction for
the hydrogen molecule and also to early results for the helium
system, which to this author's knowledge has not yet been updated.
As is now well known, the form of the Hamiltonian, derives
essentially from a large distance expansion of the
electron-electron interaction for the hydrogen molecule. This is
given by \cite{Margenau31}:
\begin{eqnarray}
H^{\prime} & = & (1/R^3)(x_1x_2+y_1y_2 - 2z_1z_2)+ \nonumber \\
&+&{3 \over 2}(1/R^4)[r_1^2z_2-r^2_2z_1\nonumber \\
&+&(2x_1x_2+2y_1y_2-3z_1z_2)(z_1-z_2)]\nonumber \\ &+&{3\over
4}(1/R^5)[r_1^2r_2^2-5r_2^2z_1^2-5r_1^2z_2^2 \nonumber \\
&-&15z_1^2z_2^2+ 2(x_1x_2+y_1y_2+4z_1z_2)^2]+ \dots \label{Heqn1}
\end{eqnarray}
In this expression $x_1,y_1,z_1$ are the Cartesian coordinates of
electron $1$ relative to its nucleus, while $x_2,y_2,z_2$ are
those of electron 2 relative to its own nucleus and the $z$ axis
is directed from one nucleus to the other. The expressions contain
the dipole-dipole interaction $\upsilon_A$(first term), the
dipole-quadrupole interaction $\upsilon_B$(second term) and the
quadrupole-quadrupole interaction $\upsilon_C$(third term), with
the first term being the well known van der Waals attraction which
is dominant for large distances $R$ - the internuclear separation.
Throughout this paper we shall be using Hartree units in which
$(e={\hbar}=a_0=1)$ and thus the absolute ground state energy
$E_0$ which we shall require later has the unit of $1/2$ Hartree.
In view of the symmetry of the various terms, it can shown
\cite{PaulingBeach35} that the above eqn(\ref{Heqn1}) is
equivalent with respect to its second-order perturbation energy to
the Hamiltonian:
\begin{eqnarray}
H^{\prime} & = & -2 [\alpha\ \xi_1\xi_2 \cos\theta_1\cos\theta_2
\nonumber
\\ &+&\beta\ \xi_1\xi_2^2 \cos\theta_1(3\cos^2\theta_2-1)\nonumber \\
&+&\gamma\ \xi_1^2\xi_2^2 (3\cos^2\theta_1-1)(3\cos^2\theta_2-1)+
\dots\ ], \label{Heqn2}
\end{eqnarray}
in polar coordinates, whereby: $\xi_{1,2} = 2 r_{1,2},
\alpha=(6)^{1\over 2}/8\ R^{-3},\beta=(30)^{1\over 2}/32\ R^{-4}\
$and$\ \gamma=(70)^{1\over 2}/128\ R^{-5}$. In this section we
shall briefly review some of the difficulties in an accurate
determination of the constants $A,B,$ and $C$. One standard
formula is the direct summation method as originally used by
Eisenchitz and London \cite{EisenchitzLondon30}, for the van der
Waals energy
:
\begin{equation}
\epsilon_A (R)  =  -{12\over R^6} \sum_{n,m} {f_{1,n}f_{1,m} \over
(1-{1\over n^2})(1-{1\over m^2})(2-{1\over n^2}-{1\over m^2})},
\label{Heqn3}
\end{equation}
where $f_{l,m}=2(E_m-E_l)|z_{l,m}|^2$ is the oscillator strength
as defined in terms of the dipole matrix elements $z_{l,m}$
between the states $(l,m)$ \cite{MahantyNinham76}.  However, the
series eqn(\ref{Heqn3}) converges badly even for the discrete
states and there are terms involving the matrix elements between
discrete/ continuum and continuum/continuum states that are
difficult to evaluate but which ultimately determine the accuracy
of the result. The value for the constant $A \approx 6.47$
originally given by Eisenchitz and London
\cite{EisenchitzLondon30} after much work testifies to the
difficulty of this approach. Nevertheless, eqn(\ref{Heqn3}) has
its appeal in that it can be recast into the form of an integral
over imaginary frequencies of the dynamical polarizabilities of
the two atoms:
\begin{equation}
\epsilon_A (R)  =  -{3\over R^6} \int_0^\infty d\xi\
\alpha_1(i\xi)\alpha_2(i\xi), \label{Heqn4}
\end{equation}
from which current theories for the van der Waals interaction for
more complex systems like He are based. These employ a combination
of linear response and time-dependent density functional theories
to obtain the polarizabilities \cite{KohnMM98}. An integral
equation is ultimately involved in the solution for
$\alpha_{1,2}(i\xi)$, generally involving various decoupling
approximations, and thereafter the integral eqn(\ref{Heqn4}) has
to be performed. To the best of this author's knowledge, none of
the theories proposed so far have been tested with the exactly
soluble case of the $H_2$ molecule \cite{DobsonDW98},
\cite{KohnMM98} providing added motivation for our present work.

Let us mention at the outset that the central difficulty of this
problem has to do with an accurate treatment of excited states,
their matrix elements with the ground state (both discrete and
continuous) and the poor convergence of the series like
eqn(\ref{Heqn3}). These are all sticky points with the modern
local density functional theories (LDFT)\cite{Nalewajski96}. It
has been identified by Slater and Kirkwood (SK) in their classic
paper \cite{SlaterKirkwood31}, that a superior method involves a
direct perturbation wavefunction ansatz of the form:
\begin{equation}
\psi({\bf r_1},{\bf r_2})  = \psi_0(r_1,r_2)[1 + \phi({\bf
r_1},{\bf r_2})], \label{Heqn5}
\end{equation}
in which $\psi_0$ is the ground state wavefunction of the
unperturbed system. The function $\phi$ (a two particle
correlation function as we shall see), satisfies the following
exact differential equation, easily derived from the
Schr{\"o}dinger equation up to first order in $\upsilon$ which can
be $ \upsilon_A,\upsilon_B$ or $\upsilon_C$ accordingly. Following
the notation of SK \cite{SlaterKirkwood31}, this is given by:
\begin{equation}
{1 \over 2}\nabla^2 \phi +  (\nabla \ln \psi_0).(\nabla \phi)-
\upsilon = 0. \label{Heqn6}
\end{equation}
SK used this equation as their basis for the treatment of the
$H_2$ and He systems. For the moment we shall concentrate on the
$H_2$ molecule, which by the substitution:
\begin{equation}
\phi={\upsilon R(\xi,\xi^{\prime})\over E_0} \label{Heqn7}
\end{equation}
leads to the differential equation (hereafter known as the SK
equation) in the case of the van der Waals interaction
$\upsilon_A$ as given by:
\begin{eqnarray}
{\partial^2 R \over \partial \xi^2} + {\partial^2 R \over \partial
\xi^{\prime 2}} &+& ({4\over \xi}-1){\partial R \over \partial
\xi}+({4\over \xi^{\prime}}-1){\partial R \over \partial
\xi^{\prime}} \nonumber \\ &-& R({1\over \xi}+{1\over
\xi^{\prime}})-{1\over 4}=0. \label{Heqn8}
\end{eqnarray}
Note that $R(\xi,\xi^{\prime})$ is strictly non-local, but it can
in principle be derived from a {\it local} density, as we shall
see. Unfortunately SK were unable to solve this inhomogeneous
(PDE) equation and resorted to various approximations, one of
which is to ignore the differential terms and assume:
\begin{equation}
R_0(\xi,\xi^{\prime})=-{1\over 4}{\xi\ \xi^\prime \over \xi +
\xi^\prime}. \label{Heqn9}
\end{equation}
They suggested that this is a good first approximation and that
subsequent approximations can be obtained by substituting this
into the differential function of eqn(\ref{Heqn8}) and iterating.
They evaluated the constant $A \approx 6.14$ using
eqn(\ref{Heqn9}), a result tabulated in the book of Pauling and
Wilson \cite{PaulingWilson35},  but unfortunately we have found
this value to be in error. The correct approximate value should be
$A \approx 6.23$. The second error in the SK paper is that their
proposed iteration method does not work. In fact it is a
non-convergent procedure, and attempts to employ it leads to
divergent results. Earlier on in this work the author has carried
out an iteration of their scheme to four orders and found the
results diverging. Nevertheless SK suggested other approximation
schemes like the ansatz $\lambda r^\nu r^{\prime \nu}$ where
$\lambda$ and $\nu$ are variational parameters (see later) which
they have found to give good results for both the $H_2$ and He
systems. We shall discuss the solution of eqn(\ref{Heqn8}) later
in the next section. Here we shall mention the best solution for
the $H_2$ problem to date. This was the widely cited paper of
Pauling and Beach (PB) \cite{PaulingBeach35}. Their solution
employs a general variational method in which the matrix elements
for the Hamiltonian were evaluated using special orthogonal
orbitals constructed for a solution to the Stark effect problem
\cite{PaulingWilson35}. They have found these orbitals to be ideal
for this problem by which all matrix elements for the interactions
$\upsilon_{A,B,C}$ can be computed accordingly.  Thereupon they
were able to set up an infinite determinant from the secular
equation which they have evaluated up to a rank of order $(26
\times 26)$ to find the energies. Their results for $A=6.49903$,
$B=124.399$ and $C=1135.21$ (accurate up to the last decimal)
remain the most accurate to date, until our present work and is
very impressive for 1935. However they cautioned that their
``treatment has not led to an exact solution" owing to the
uncertain nature of their variational method and the convergence
properties of their wavefunctions \cite{PaulingWilson35}. In
addition their method does not yield the expansion coefficients
for the wavefunctions, needed to ensure normalisability and
thereby obtain the normalisation constant $C(R)$. Furthermore
their procedure cannot obtain the perturbation in the charge
density which will be of interest to us here.  The reader is
referred to their earlier paper for information
\cite{PaulingBeach35}, many of whose details are now merely of
historical interests. Neverthless, they have identified their
method as identical with and is a more generalised variational
form of that used by Hass{\'e} \cite{Hasse31}, who also was the
first to treat the $H_2$ and He problems. In the next section we
shall discuss the exact solution that SK had failed to obtain for
eqn(\ref{Heqn8}). We should note that the power of their method
lies in the ability to employ the interaction function itself to
project out {\it all} relevant components of the excited states
for second order perturbation theory, into the two particle
correlation function $R(r,r^\prime)$. This then has a concise form
satisfying an inhomogeneous PDE eqn(\ref{Heqn8}). This method was
later generalised to higher order perturbation theory by Dalgarno
and Lewis \cite{DalgarnoLewis55} and Schwartz \cite{Schwartz59}
see also Schiff \cite{Schiff68}.
\section{Solution of the SK equation}
\label{Exact} Having set the background to the hydrogen problem we
shall next discuss the solution of the SK equation
eqn(\ref{Heqn8}). This is a two dimensional inhomogeneous (PDE)
and we must first start by discussing the appropriate boundary
conditions. This boundary value problem is unusual in that it is
not of the standard Dirichlet or Neumann type as is common in
electrostatics \cite{Jackson75}. In fact there are no particular
{\it a prior} boundary values apart from the requirement for the
normalisability of the wavefunction, and special values such as
$R(0,0)$ are determined only after the solution is obtained.
Nevertheless an analogous Green's function integral equation
method which is exact is known to exist due to the work of Levi
\cite{CourantHilbert}. This is of no interest to us here, so that
we shall merely outline the method in an appendix, but it may be
useful for making connections with other integral equation methods
of treating the problem, such as via linear response theory
\cite{DobsonDW98}, \cite{KohnMM98}.

The method we shall use to solve eqn(\ref{Heqn8}) is gained from
experience in solving the two sphere problem of classical
electrostatics, \cite{Choy98a},\cite{Choy98b}.  As we shall see
the convergence of this problem by our method is superior to the
electrostatic case of two spheres, which required nearly 200 terms
for convergence to only two decimal places\cite{Choy98a}.  We
begin by expanding the two particle correlation function in the
following ansatz in terms of orthogonal polynomials:
\begin{equation}
R(\xi,\xi^{\prime})= \sum_{l,n} a_{l,n} H_l(\xi)
H_n(\xi^{\prime}), \label{Heqn10}
\end{equation}
where the functions $H_n(x) = L^3_{n+1}(x)$ are defined in terms
of associated Laguerre polynomials \cite{comment1}. Inserting this
into the SK eqn(\ref{Heqn8}), it reduces to the form:
\begin{equation}
\sum_{l,n} a_{l,n} \Bigl[ {(l-1)\over \xi} + {(n-1)\over
\xi^{\prime}} \Bigr ] H_l(\xi) H_n(\xi^{\prime}) = -{1\over 4},
\label{Heqn11}
\end{equation}
when use is made of the equation for Laguerre polynomials. Upon
multiplying both sides of eqn(\ref{Heqn11}) by:
\begin{equation}
\xi^4 e^{-\xi} H_m(\xi) \xi^{\prime 4} e^{-\xi^\prime}
H_s(\xi^\prime)\label{Heqn12}
\end{equation} and then integrating, with the use of the properties
of the Laguerre polynomials,  we derive the following infinite set
of linear equations for the $a_{m,s}$:
\begin{eqnarray}
&-& a_{m,s}(2 s g_m q_s + 2 m g_s q_m) + a_{m,s-1} (s+1) g_m
q_{s-1} \nonumber \\ &+& a_{m,s+1} (s-1) g_m q_{s+1} + a_{m-1,s}
(m+1) g_s q_{m-1} \nonumber \\  &+& a_{m+1,s} (m-1) g_s q_{m+1}=
\Delta_{m,s} . \label{Heqn13}
\end{eqnarray}
where:
\begin{eqnarray}
g_s &=& {1\over 144} {(s-1)(s+1)! \over (s-2)!} \ , q_s =
{(s+1)!\over (s-2)!},  \nonumber \\ & & \Delta_{m,s} =
\delta_{m,2}
\delta_{s,2}-\delta_{m,3}\delta_{s,2}-\delta_{m,2}\delta_{s,3}
+\delta_{m,3}\delta_{s,3}. \label{Heqn14}
\end{eqnarray}
This set of equations is readily solved symbolically using
Mathematica version 3.1 on a PC.  We shall present the results in
the next subsection.  Here we shall obtain the form of the energy
expressions. Firstly, as was shown first by SK
\cite{SlaterKirkwood31}, for an arbitrary correlation function
${\tilde R}$ not necessarily satisfying eqn(\ref{Heqn8}), the
energy expression is given by:
\begin{eqnarray}
\epsilon_A &=& {1 \over (16 \pi E_0)} \int_0^\infty
d{\xi}\int_0^\infty d{\xi^\prime} \nonumber \\ & &
\upsilon_A^2(\xi,\xi^\prime){\tilde R}(\xi,\xi^\prime)\Bigl(
{1\over 4}-{\cal L}[{\tilde R}(\xi,\xi^\prime)]\Bigr ),
\label{Heqn15}
\end{eqnarray}
where ${\cal L}$ is the differential operator given by the LHS of
eqn(\ref{Heqn8}). This form is particuarly useful when we look at
density functional theories afterwards. At this point we shall
merely mention that the change necessary for calculating the
dipole-quadrupole  energy $\epsilon_B$ and the
quadrupole-quadrupole energy $\epsilon_C$ are the form of
$\upsilon$ which becomes $\upsilon_B$ and $\upsilon_C$
respectively. The form of $R$ also changes which we shall call
$R_B$ and $R_C$, (subscripts being only used when there is a need
to avoid confusion) and so are the operators ${\cal L}_B$ and
${\cal L}_C$ . They can be obtained from the appropriate SK
equations and are given by:
\begin{eqnarray}
{\partial^2 R_B \over \partial \xi^2} + {\partial^2 R_B \over
\partial \xi^{\prime 2}} &+& ({4\over \xi}-1){\partial R_B \over
\partial \xi}+({6\over \xi^{\prime}}-1){\partial R_B \over \partial
\xi^{\prime}} \nonumber \\ &-& R_B({1\over \xi}+{2\over
\xi^{\prime}})-{1\over 4}=0, \label{Heqn16}
\end{eqnarray}
and
\begin{eqnarray}
{\partial^2 R_C \over \partial \xi^2} + {\partial^2 R_C \over
\partial \xi^{\prime 2}} &+& ({6\over \xi}-1){\partial R_C \over
\partial \xi}+({6\over \xi^{\prime}}-1){\partial R_C \over \partial
\xi^{\prime}} \nonumber \\ &-& 2 R_C({1\over \xi}+{1\over
\xi^{\prime}})-{1\over 4}=0. \label{Heqn17}
\end{eqnarray}
They can be solved by the same method as for $R_A$, the
expressions for the expansion eqn(\ref{Heqn10}) now being:
\begin{equation}
R_B(\xi,\xi^{\prime})= \sum_{l,n} b_{l,n} H_l(\xi)
G_n(\xi^{\prime}), \label{Heqn18}
\end{equation}
 and
\begin{equation}
R_C(\xi,\xi^{\prime})= \sum_{l,n} c_{l,n} G_l(\xi)
G_n(\xi^{\prime}), \label{Heqn19}
\end{equation}
where $G_n(x)=L^5_{n+2}(x)$ is a higher order associated Laguerre
polynomial. We note that for the exact solution ${\cal L}[
R(\xi,\xi^\prime)] = 0$ from which the energy expressions are
easily obtained in terms of the first few expansion coefficients.
We shall collect the formulas for the various energy constants in
terms of these coefficients as:
\begin{eqnarray}
A&=& - 12 [a_{2,2}-a_{2,3}-a_{3,2}+a_{3,3}], \nonumber \\ B&=& -
270 [b_{2,3}-b_{2,4}-b_{3,3}+b_{3,4}], \nonumber \\ C&=& - 2835
[c_{3,3}-c_{3,4}-c_{4,3}+c_{4,4}]. \label{Heqn20}
\end{eqnarray}
Thus the energy constants can be determined to any desired
accuracy using symbolic manipulation codes such as by Mathematica
version 3.1 which yields the $a,b,c$ coefficients as exact
fractions.

In order to check the definite convergence of the wavefunctions, a
point of concern for Pauling and Beach \cite{PaulingBeach35}, we
have also computed the normalization constants for the
wavefunctions in eqn(\ref{Heqn5}). These are given by the
integrals:
\begin{equation}
C_i(R) = \pi^2 + {1\over E_0^2} \int d{\bf r} \int d{\bf r}^\prime
\psi_0^2 {\upsilon_i^2 R_i^2}, \label{Heqn21}
\end{equation}
where $i=A,B,C$ respectively. More appropriate expressions are
given in terms of the constants $D_i$ where we have factored out
the distance dependence $R$:
\begin{eqnarray}
C_A &=& \pi^2 (1 + {D_A \over R^6}), \nonumber \\ C_B &=& \pi^2 (1
+ {D_B \over R^8}), \nonumber \\ C_C &=& \pi^2 (1 + {D_C \over
R^{10}}). \label{Heqn22}
\end{eqnarray}
The values of $D_i$ can thus be obtained in a series in terms of
the $a,b,c$ coefficients and computed to any desired accuracy. Of
particular interest to us is a calculation of the density. This
can be obtained by direct partial integration of the wavefunction
in eqn(\ref{Heqn5}). Since we are only interested in the density
perturbation, by subtracting the unperturbed density $\rho_0=\pi
e^{-\xi}$ for the ground state hydrogen atom, we have:
\begin{eqnarray}
{\delta\rho(\xi,\theta)\over \rho_0} &=& {\rho(\xi,\theta)-\rho_0
\over \rho_0} \nonumber \\ &=& \int_{0}^{\pi} d\theta^{\prime}
\sin\theta^{\prime} \int_{0}^{\infty} d\xi^{\prime}
e^{-\xi^{\prime}} \xi^{\prime 2} \nonumber \\ & &
\upsilon^2(\xi,\xi^\prime,\theta,\theta^\prime)R^2(\xi,\xi^\prime).
\label{Heqn23}
\end{eqnarray}
We shall redefine alternative functions $f_i(\xi)$ which we shall
call ``densities" and they are the main focus in this paper:
\begin{eqnarray}
{\delta\rho_A(\xi,\theta)\over \rho_0} &=& {1\over 4 R^6} \xi^2
\cos^2\theta f_A(\xi) \nonumber \\
{\delta\rho_B^{(1)}(\xi,\theta)\over \rho_0} &=& {3\over 16 R^8}
\xi^2 \cos^2\theta f_B^{(1)}(\xi) \nonumber \\
{\delta\rho_B^{(2)}(\xi,\theta)\over \rho_0} &=& {5\over 64 R^8}
\xi^4 (3\cos^2\theta-1)^2 f_B^{(2)}(\xi) \nonumber \\
{\delta\rho_C(\xi,\theta)\over \rho_0} &=& {7\over 256 R^{10}}
\xi^4 (3\cos^2\theta-1)^2 f_C(\xi). \label{Heqn24}
\end{eqnarray}
Note that there are two densities for B, since the
dipole-quadrupole interaction is asymmetric. All these densities
are readily computed from the appropriate wavefunction
coefficients. For example we have:
\begin{equation}
f_A(\xi)=\sum_{n,m}\alpha_{n,m} H_n(\xi)H_m(\xi)\  {\rm etc.},
\label{Heqn24b}
\end{equation}
where:
\begin{eqnarray}
\alpha_{n,m}=\sum_l a_{n,l}[2l q_l
a_{m,l}&-&(l+1)q_{l-1}a_{m,l-1}\nonumber
\\&-&(l-1)q_{l+1}a_{m,l+1}], \label{Heqn24c}
\end{eqnarray}
and so on. Their results will be given in the next subsection.
\subsection{Exact results} \label{subexact} The infinite set of linear
equations such as eqn(\ref{Heqn13}) is truncated at each order and
the solution for the coefficients $a,b,c$ are solved by the use of
Mathematica on a PC accordingly.  It is remarkable that the
results are so fast converging unlike that for the two spheres
problem in electrostatics \cite{Choy98a}. Previously for the two
spheres we have found the need to export the codes to a Silicon
Graphics workstation running Mathematica, as expansions up to the
order of 200 coefficients are necessary before we could obtain
convergence to 2 decimal places. The present codes run readily on
a PC and as the following Table \ref{table1} shows they converge
rapidly. The superiority of convergence of our approach versus
other methods \cite{EisenchitzLondon30} is obvious from the
results shown in this table. Our convergence rate is even better
than the calculations of PB \cite{PaulingBeach35}.

\begin{table}
\caption{Computed constants A and $D_A$} \label{table1}
\begin{tabular}{ccc}
order of truncation  & A & $D_A$ \\ \hline \\ 1 & 6 & 6 \\
2&6.22222222$\dots$& 6.61728395$\dots$ \\ 3&6.46153846$\dots$&
6.88757396$\dots$ \\4&6.48214285$\dots$& 7.40242346$\dots$ \\
5&6.49844398$\dots$&7.40024688$\dots$\\6&6.49900257$\dots$&7.39872679$\dots$
\\7&6.49902535$\dots$&7.39863094$\dots$\\8&6.49902659$\dots$&7.39862559$\dots$
\\9&6.49902669$\dots$&7.39862525$\dots$\\10&6.49902670$\dots$&7.39862522$\dots$
\end{tabular}
\end{table}
\begin{table}
\caption{Computed constants B and $D_B$} \label{table2}
\begin{tabular}{ccc}
order of truncation  & B & $D_B$ \\ \hline \\
1&115.71428571$\dots$&24.79591836$\dots$\\ 2&118.96875&
26.74423828$\dots$\\ 3&124.26672692$\dots$&30.14754412$\dots$
\\4&124.39502505$\dots$&30.12987113$\dots$\\
5&124.39891831$\dots$&30.12699933$\dots$\\
6&124.39907397$\dots$&30.12683881$\dots$\\
7&124.39908277$\dots$&30.12682990$\dots$\\
8&124.39908349$\dots$&30.12682930$\dots$\\
9&124.39908357$\dots$&30.12682924$\dots$\\
10&124.39908358$\dots$&30.12682924$\dots$
\end{tabular}
\end{table}

Here we see that at truncation order 10, where there are about 100
coefficients, the smallest being $a_{10,10}$, we have achieved
convergence to at least seven decimal places. Our results are in
exact agreement with Pauling and Beach \cite{PaulingBeach35},
indicating that they have indeed found the exact energy
variationally. In particular the normalisation constants converge
to the same accuracy indicating that the wavefunctions are
normalisable and thus well behaved. In the following
Fig.\ref{fig1} we shall show the computed density function
$f_A(\xi)$.
\begin{figure}[htb]
\centerline{\epsfxsize=3.3in \epsffile{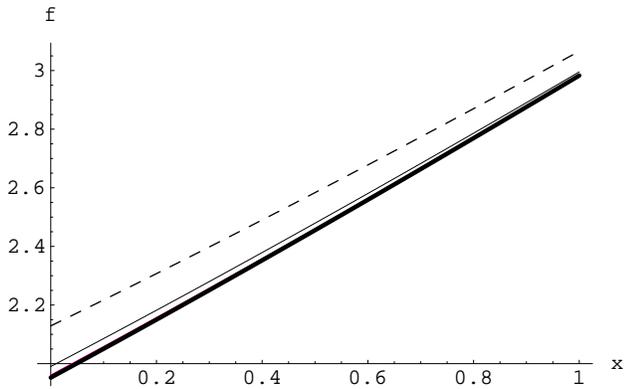}}
\caption{Computed density function $f_A(\xi)$ for successive
truncations from order 5 (dashes). Note the convergence in the
successive plots towards the solid curve.}\label{fig1}
\end{figure}
\begin{table}
\caption{Computed constants C and $D_C$} \label{table3}
\begin{tabular}{ccc}
order of truncation  & C & $D_C$ \\ \hline \\ 1
&1063.125&199.3359375\\2&1132.610294117$\dots$&238.583207179$\dots$
\\3&1135.107421875&238.686733245$\dots$\\
4&1135.208820466$\dots$&238.645331783$\dots$ \\
5&1135.213725627$\dots$&238.641796376$\dots$\\
6&1135.214015982$\dots$&238.641543683$\dots$\\
7&1135.214037581$\dots$&238.641524775$\dots$\\
8&1135.214039617$\dots$&238.641523160$\dots$\\
9&1135.214039858$\dots$&238.641522996$\dots$\\
10&1135.214039892$\dots$&238.641522976$\dots$
\end{tabular}
\end{table}
It is to be noted that the large distance behaviour in the density
dictates the subsequent accuracy in the energy calculation. We
show the density at various orders in Fig.\ref{fig2}.
\begin{figure}[htb]
\centerline{\epsfxsize=3.3in \epsffile{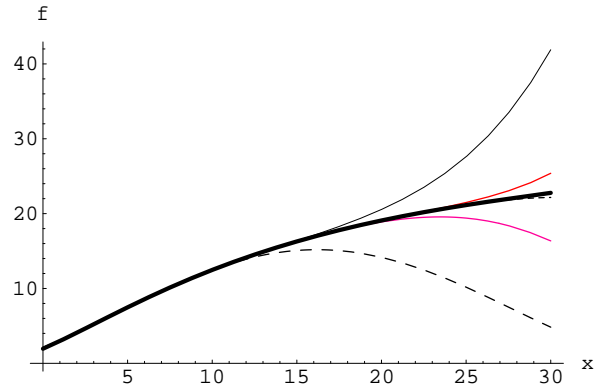}}
\caption{Computed density function $f_A(\xi)$ at larger distances
for successive truncations. Order 5 is the bottom curve (dashes)
while order six is the top curve (line). The curves have
alternative curvatures for odd and even orders of truncation. Note
the convergence to the middle plots at higher orders and the
difference with the scale of Fig.\ref{fig1}. } \label{fig2}
\end{figure}

In a similar way we have computed the B and C energy constants in
eqn(\ref{Heqn1a}), as well as their respective normalisation
constants $D_B$ and $D_C$ respectively. These are shown in Tables
\ref{table1} and \ref{table2}. Note that the numbers are tabulated
as decimals for convenience, but they are exact fractions as given
by Mathematica. As such some of the numbers terminate as decimals,
whereas the others have a finite period, most of which are much
longer than the 8 or 9 decimals shown. For completeness of the
results we have plotted all the various density functions
calculated from the exact wavefunction coefficients.
\begin{figure}[htb]
\centerline{\epsfxsize=3.3in \epsffile{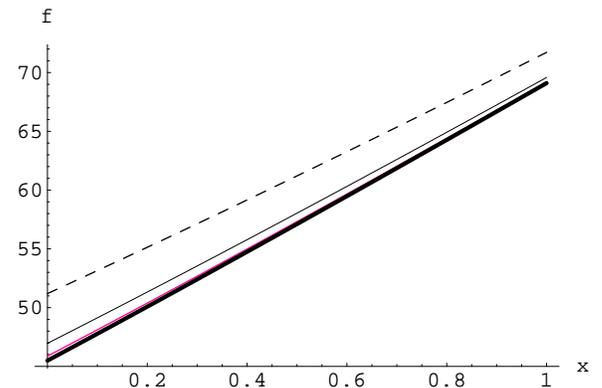}}
\caption{Computed density function $f_B^{(1)}(\xi)$ for successive
truncations from order 5 (dashes), as in
Fig.\ref{fig1}.}\label{fig3a}
\end{figure}
\begin{figure}[htb]
\centerline{\epsfxsize=3.3in \epsffile{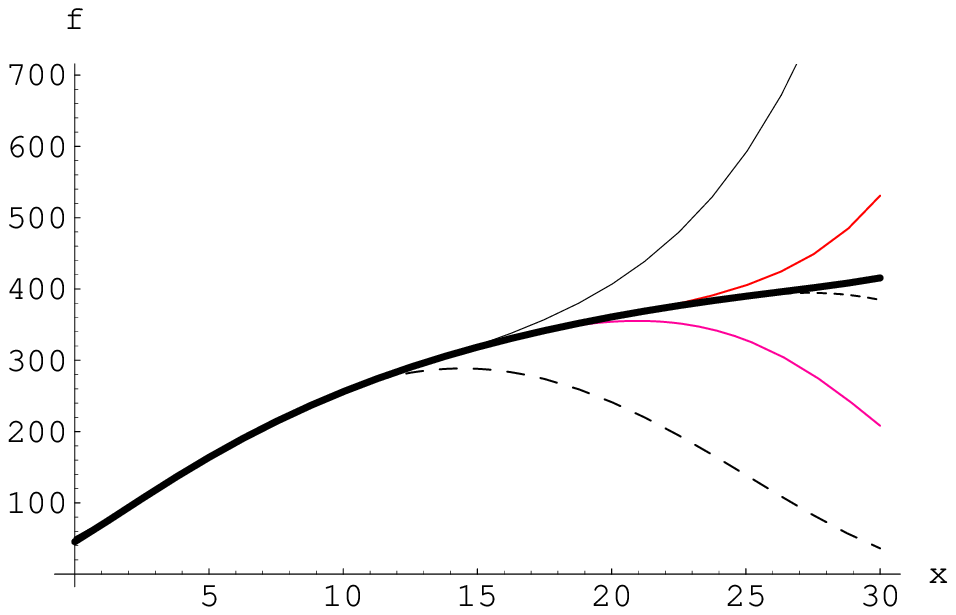}}
\caption{Computed density function $f_B^{(1)}(\xi)$ at larger
distances, for successive truncations from order 5 (bottom curve),
as in Fig.\ref{fig2}. } \label{fig3b}
\end{figure}
\begin{figure}[htb]
\centerline{\epsfxsize=3.3in \epsffile{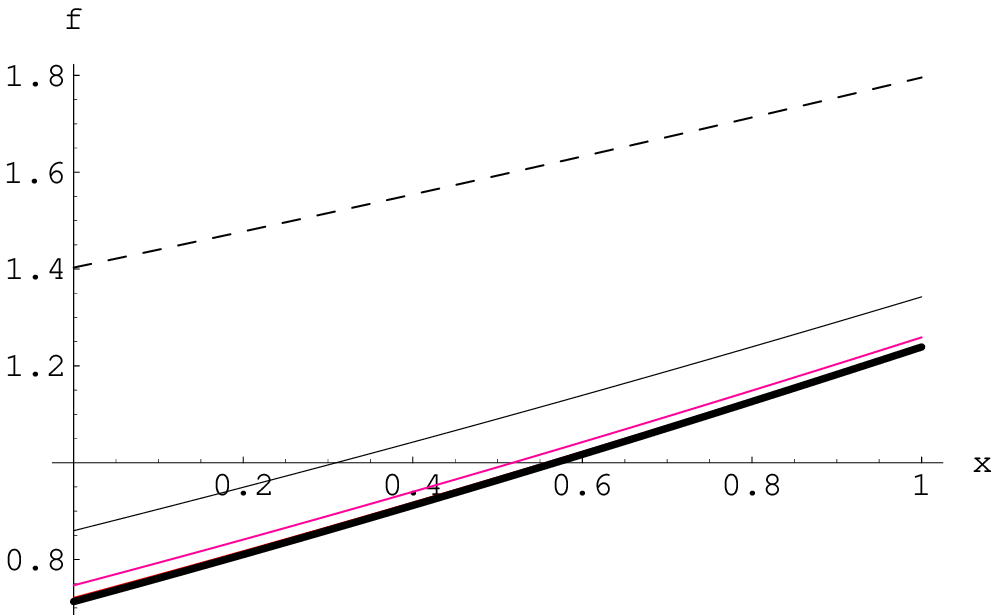}}
\caption{Computed density function $f_B^{(2)}(\xi)$ for successive
truncations from order 5 (dashes), as in Fig.\ref{fig3a}.
}\label{fig3c}
\end{figure}
\begin{figure}[htb]
\centerline{\epsfxsize=3.3in \epsffile{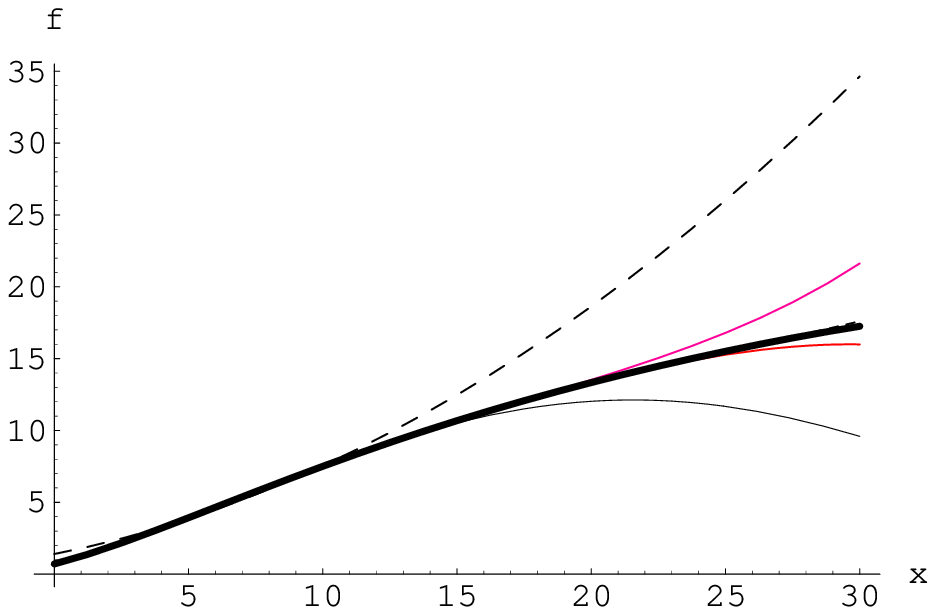}}
\caption{Computed density function $f_B^{(2)}(\xi)$ at larger
distances for successive truncations from order 5 (bottom curve),
as in Fig.\ref{fig3b} } \label{fig3d}
\end{figure}
\begin{figure}[htb]
\centerline{\epsfxsize=3.3in \epsffile{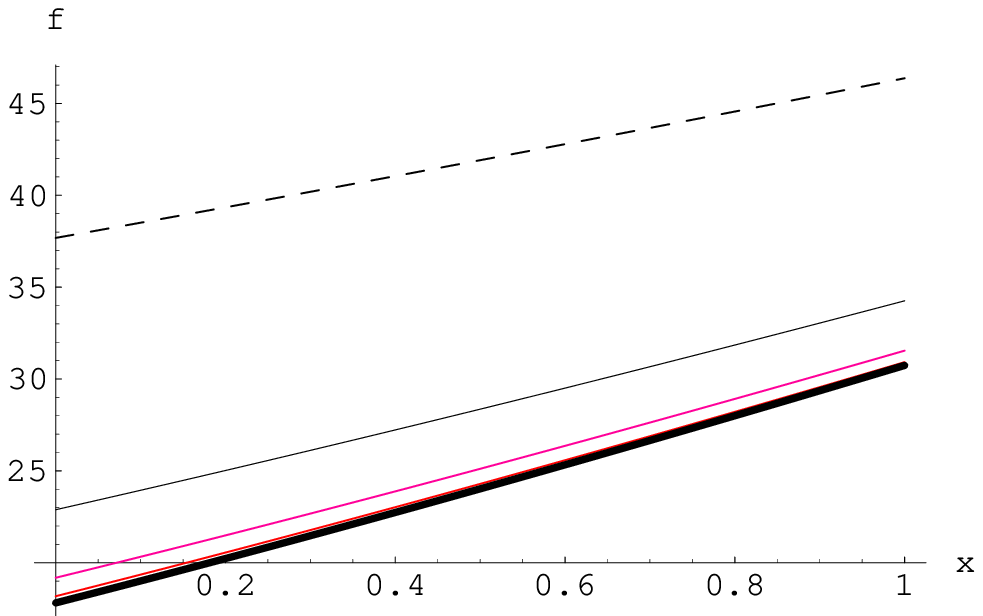}}
\caption{Computed density function $f_C(\xi)$ for successive
truncations from order 5 (dashes), as in Fig.\ref{fig3c}
.}\label{fig4a}
\end{figure}
\begin{figure}[htb]
\centerline{\epsfxsize=3.3in \epsffile{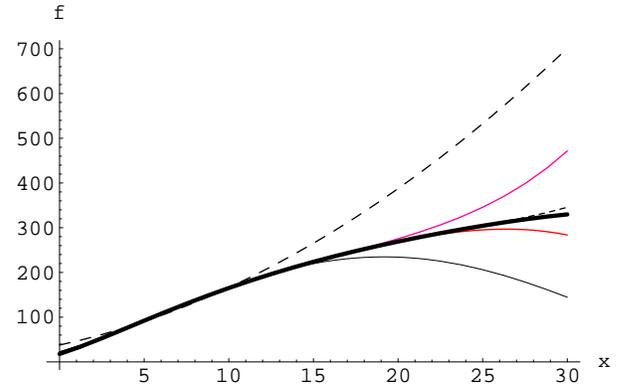}}
\caption{Computed density function $f_C(\xi)$  at larger distances
for successive truncations from order 5 (bottom curve), as in
Fig.\ref{fig3d} } \label{fig4b}
\end{figure}
Note that for shorter distances in figures
\ref{fig1},\ref{fig3a},\ref{fig3c} and \ref{fig4a}, the
convergence proceeds monotonically from the top curve to the
bottom.  These plots are nearly straight, but for larger distances
they have alternative curvatures for odd and even truncation
orders. These observations furnish interesting approximation
methods, which will not be investigated here.  In the next
section, we shall use the results obtained so far to derive an
exact local density functional theory for this system.
\section{An exact energy local density functional}
\label{DFT} Current thinking in density functional theory (DFT)
approaches the problem of dispersion forces from eqn(\ref{Heqn4}),
as mentioned earlier, via time-dependent generalisations of DFT.
It is interesting to note that the method of Kohn et al
\cite{KohnMM98} has yielded results in agreement with the best
theoretical value for He up to two decimal places, in spite of a
7\% error in the completeness sum rule in their calculations.
However it is difficult from these theories to develop systematic
improvement methods and would require considerable expertise in
time-dependent density functional methods.  The question arises a
long ago in a paper pointed out by Lieb \cite{Lieb83}, that the
universal Hohenberg-Kohn (HK) functional $I[\rho]$
\cite{HohenbergKohn} has ``hidden complexities" with respect to
the van der Waals interaction. The dynamical dipole fluctuation
properties leading to the latter ``is somehow built into
$I[\rho]$, but an explicit form of $I[\rho]$ that will produce
this effect has yet to be displayed." This is in spite of a very
general proof for the universal character of van der Waals forces
for Coulomb systems \cite{LiebThirring86}. We shall answer this
question some way for the $H_2$ molecule. It is noteworthy that
perhaps as a result of these remarks, the search for an accurate
static local density functional theory for dispersion forces has
more or less been abandoned.  In this section we shall demonstrate
that for the $H_2$ molecule system at hand we can formulate an
exact static DFT. We start from the observation that the two
particle correlation $R(r,r^\prime)$ {\it is} a functional of the
density $f_A$.  In eqn(\ref{Heqn24b}) {\it if} the quantities
$\alpha_{n,m}$ are fixed, thereby fixing the density, then
eqn(\ref{Heqn24c}) {\it in principle} can be inverted to obtain
the coefficients $a_{n,m}$ thereby determining $R(r,r^\prime)$.
Therefore upon substituting this $R[\alpha_{n,m}]$ into the energy
functional eqn(\ref{Heqn15}), then a variation of $\epsilon_A$
with respect to $\alpha_{n,m}$ would yield the exact ground state
energy. This corresponds to the constraint search algorithm of
Levy \cite{Levy79}, but note that this is {\it not} the HK
functional as it is specific to this problem. Formally we can
write this as:
\begin{eqnarray}
\epsilon_A &=& {1 \over (16 \pi E_0)} \int_0^\infty d{\bf
\xi}\int_0^\infty d{\bf \xi^\prime} \nonumber \\ & &
\upsilon_A^2(\xi,\xi^\prime){\tilde R}[f_A(\xi)]\Bigl( {1\over
4}-{\cal L}[{\tilde R}[f_A(\xi)]]\Bigr ). \label{Heqn25}
\end{eqnarray}
Then the ground state energy and density can be obtained from:
\begin{equation}
{\delta \epsilon_A[f_A(\xi)] \over \delta f_A(\xi)} = 0 .
\label{Heqn26}
\end{equation}
However eqn(\ref{Heqn24c}) is not the most convenient to use. Its
inversion corresponds to a non-linear programming problem with
many solutions. This complexity came from our choice of orbitals
which for the exact solution fixes: ${\cal L}[R[f_A(\xi)]]=0$. An
alternative choice of orbitals can be made which then fixes
${\tilde R}[f_A(\xi)]$ but now ${\cal L}[{\tilde
R}[f_A(\xi)]]\not=0$. Nevertheless eqn(\ref{Heqn26}) will still
yield the exact ground state energy via eqn(\ref{Heqn25}) which is
the essence of our DFT.

A convenient choice of orbitals is determined from the density
expression:
\begin{equation}
f_A(\xi)= \int_0^\infty d\xi^\prime e^{-\xi^\prime}\xi^{\prime
4}{\tilde R}^2_A(\xi,\xi^\prime) . \label{Heqn27}
\end{equation}
It can be easily seen that the choice of orbitals
$\phi_n(\xi)=L^4_n(\xi)$ such that:
\begin{equation}
f_A(\xi)=\sum_{n,l}{\tilde \alpha}_{n,l} \phi_n(\xi)\phi_l(\xi)\
{\rm etc.}, \label{Heqn28}
\end{equation}
provide a much simpler form for the density whereupon:
\begin{equation}
{\tilde R}_A(\xi,\xi^\prime)= \sum_{n,m} {\tilde a}_{n,m}
\phi_n(\xi) \phi_m(\xi^\prime), \label{Heqn29}
\end{equation}
where:
\begin{equation}
{\tilde \alpha}_{n,l}=\sum_m \omega_m {\tilde a}_{n,m} {\tilde
a}_{l,m} ; \label{Heqn30}
\end{equation}
in which $\omega_m=m!/(m-4)!$, as appropriate for these orbitals.
The variation of the density $f_A$ can now be effected by directly
varying the coefficients ${\tilde a}_{n,m}$. The latter can be
easily computed from eqn(\ref{Heqn26}), which can be carried out
symbolically as well. The integrals required throughout the
calculation can also be computed in closed form, facilitated by
the symbolic integration capabilities of Mathematica. Our results
are tabulated in Table \ref{table4}. The results are in exact
agreement with Table \ref{table1}, since as we have noted earlier
the output are exact numeric fractions that can be compared order
by order with the results of the subsection \ref{subexact}. The
integrals involved are somewhat more lengthy here so that we have
not extended the calculations beyond the 6th order.  In the
following figures, we have plotted the density functions which are
again in exact agreement with the previous figures \ref{fig1} and
\ref{fig2}.
\begin{figure}[htb]
\centerline{\epsfxsize=3.3in \epsffile{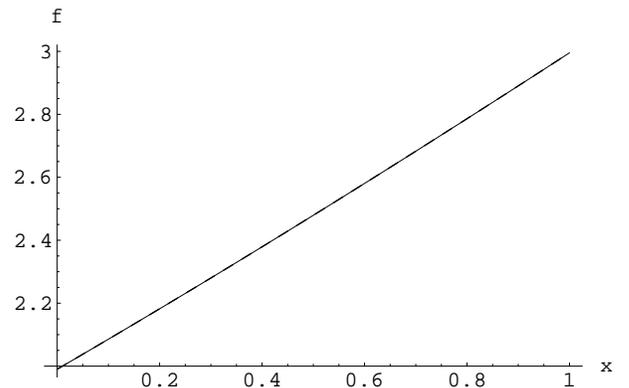}}
\caption{Computed density function $f_A(\xi)$ truncated at order
six, the dotted curve is our exact DFT result and the solid curve
is the exact result of the previous section. Both curves are
identical.}\label{fig5a}
\end{figure}
\begin{table}
\caption{Computed constant A using our DFT eqn(\ref{Heqn26}}
\label{table4}
\begin{tabular}{cc}
order of truncation  & A  \\ \hline \\ 1 & 6 \\
3&6.46153846$\dots$\\4&6.48214285$\dots$\\
5&6.49844398$\dots$\\6&6.49900257$\dots$
\end{tabular}
\end{table}
\begin{figure}[htb]
\centerline{\epsfxsize=3.3in \epsffile{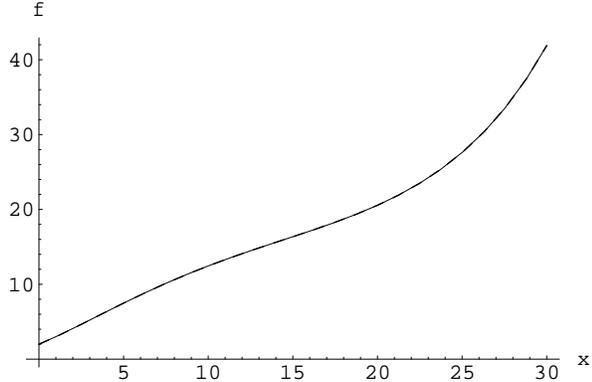}}
\caption{Computed density function $f_A(\xi)$ for truncations at
order 6 for larger distances. The dash curve is our exact DFT
result and the solid curve is the exact result of the previous
section, both curves are seen to be identical.} \label{fig5b}
\end{figure}
We can show in the same way using the exact DFT for the other
energy constants B and C that they yield identical results as
subsection \ref{subexact}. The convenient orbitals for these other
density expansions are easily seen to be: $\phi_n(\xi)
\chi_m(\xi)$ for $f_B^{(1)}(\xi)$ and $\chi_n(\xi) \chi_m(\xi)$
for $f_B^{(2)}(\xi)$ and $f_C^{(2)}(\xi)$ accordingly, where
$\chi_n(\xi)=L^6_n(\xi)$.  We found these results to be very
instructive from the viewpoint of DFT. In particular approximate
densities can be developed. For example the form:
\begin{equation}
f_A(\xi)= {\rm Const}\quad \xi^{2\nu}, \label{Heqn31}
\end{equation}
with Const $=\lambda^2 (4+2\nu)!$ follows from the SK ansatz for
$R(r,r^\prime)=\lambda r^\nu r^{\prime \nu}$. The variational
results using this approximation (which have to be computed
numerically) give two significant figures accuracy except for the
case of $\epsilon_B$, see Table \ref{table5}.
\begin{table}
\caption{Computed constants A,B and C using eqn(\ref{Heqn31})}
\label{table5} \begin{tabular}{ccc} A  & B & C \\ \hline \\
6.48965 & 116.795 & 1134.71
\end{tabular}
\end{table}The following figures compare the approximate with
the exact densities. We shall not discuss these approximations
here as detail investigations will require further work. In the
next section we shall discuss if our results could be extended as
approximate methods for more complex systems for which no exact
solutions are known.
\begin{figure}[htb]
\centerline{\epsfxsize=3.3in \epsffile{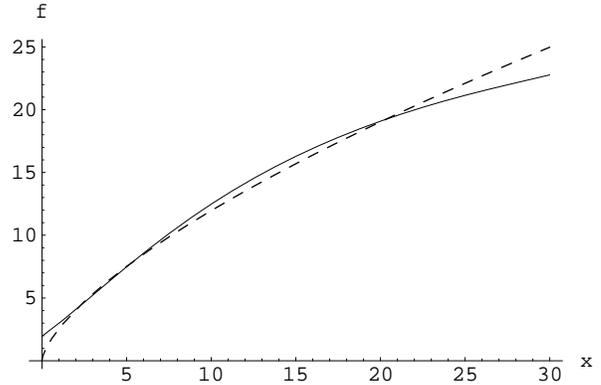}}
\caption{Approximate density function $f_A(\xi)$ (dashed)
calculated from eqn(\ref{Heqn31}) compared with the exact
one.}\label{fig6a}
\end{figure}
\begin{figure}[htb]
\centerline{\epsfxsize=3.3in \epsffile{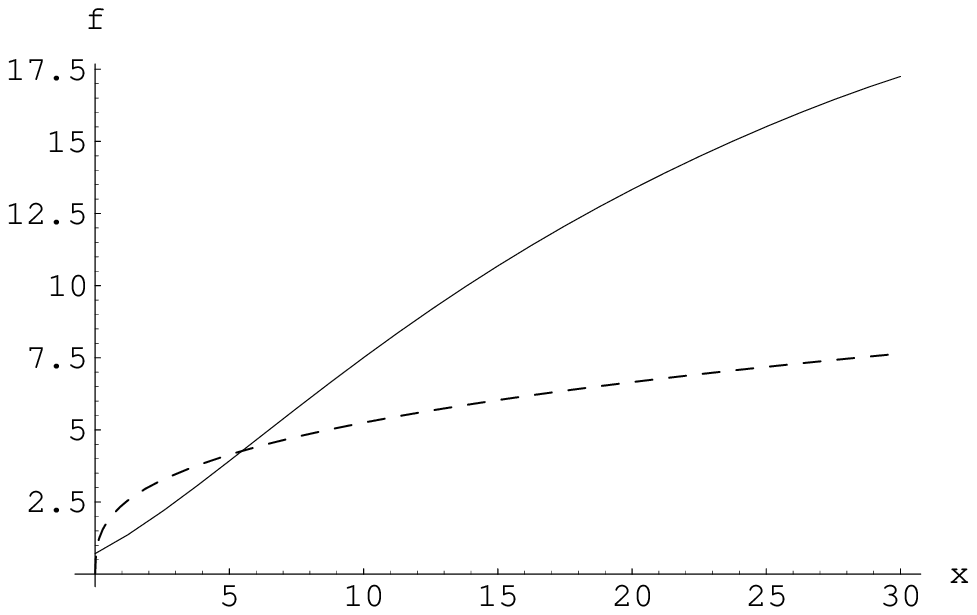}}
\caption{Approximate density function $f_B^{2}(\xi)$ (dashed)
calculated from eqn(\ref{Heqn31}) compared with the exact one.
Note that this poor density also yields a poor energy
constant.}\label{fig6b}
\end{figure}
\begin{figure}[htb]
\centerline{\epsfxsize=3.3in \epsffile{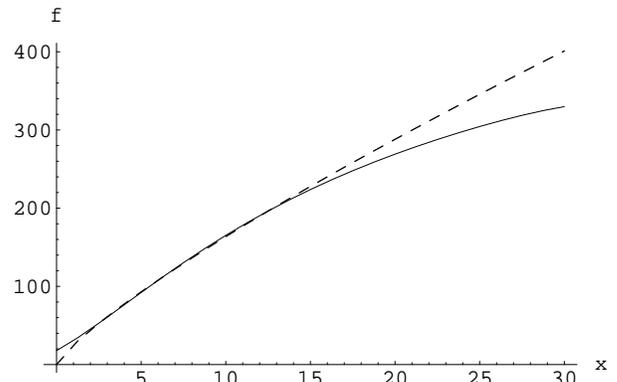}}
\caption{Approximate density function $f_C(\xi)$ (dashed)
calculated from eqn(\ref{Heqn31}) compared with the exact one.
Note that this good density also yields a good energy
constant.}\label{fig6c}
\end{figure}

\section{Hydrogenic systems and helium}
\label{Helium} For hydrogenic atoms for which we can replace the
core charge from $Z=1$ by $Z^\prime$ say, the modifications of our
method is quite straightforward. A first principles calculation
however is a different matter. We shall only discuss the van der
Waals energy $\epsilon_A$ from hereon. This can be seen by
considering the helium problem. The interaction energy contains a
sum of interactions taken from all possible pairs of electrons
between the two atoms:
\begin{equation}
\upsilon = \sum_{j=1}^N \sum_{k=1}^N \upsilon_{j,k},
\label{Heqn32}
\end{equation}
where $N$ is the number of electrons on each atom and:
\begin{equation}
\upsilon_{j,k}= (1/R^3)(x_jx_k+y_jy_k - 2z_jz_k). \label{Heqn33}
\end{equation}
The two particle correlation function eqn(\ref{Heqn7}) now breaks
up into a sum of pairs:
\begin{equation}
\phi({\bf r}_j,{\bf r}_k) = {1\over E_0} \sum_{j=1}^N \sum_{k=1}^N
\upsilon_{j,k}R(r_j,r_k), \label{Heqn34}
\end{equation}
and the SK eqn(\ref{Heqn8}) now becomes a set of $N^2$ equations:
\begin{eqnarray}
{\partial^2 R \over \partial \xi_j^2} &+& {\partial^2 R \over
\partial \xi_k} + ({4\over \xi_j}+2{\partial \ln \psi_0 \over
\partial \xi_j}){\partial R \over \partial \xi_j}+({4\over \xi_k}+2{\partial
\ln \psi_0 \over \partial \xi_k}){\partial R \over \partial \xi_k}
\nonumber \\ &+& 2 R({1\over \xi_j}{\partial \ln \psi_0 \over
\partial \xi_j}+{1\over \xi_k}{\partial \ln \psi_0 \over
\partial \xi_k})-{1\over
4}=0. \label{Heqn35}
\end{eqnarray}
These equations are now coupled, since in general $\psi_0$ is a
many-body wavefunction. Hence the problem is in general insoluble.
Nevertheless, the situation in which $\psi_0$ is given in an LDA
approximation as a product of Kohn-Sham (KS) orbitals
\cite{KohnSham65} is greatly simplified and should form the basis
of an LDA approach as we shall see . We have learnt from the
hydrogen problem that the correlation function $R(r_j,r_k)$ and
the density $f(r_j)$ are closely connected with these orbitals.
For hydrogenic atoms in which the core can be considered as a
closed shell, then the following approximation:
\begin{equation}
{\partial \ln \psi_0 \over \partial \xi_{j,k}} \approx
-Z^{\prime}/2, \label{Heqn36}
\end{equation}
where $Z^{\prime}<1$ can be made without a significant lost of
accuracy. In this case eqn(\ref{Heqn35}) reduces to a single
equation and is amenable to analytical treatment as for hydrogen.
However from the form of the energy, which is additive in terms:
\begin{equation}
\epsilon=\sum_{j=1}^N \sum_{k=1}^N \epsilon_{j,k}, \label{Heqn37}
\end{equation}
where:
\begin{eqnarray}
\epsilon_{j,k} &=& {1\over E_0} {\int \upsilon^2_{j,k} {\tilde
R}_{j,k}(1-{\cal L}_{j,k}[{\tilde R}_{j,k}]) \psi_0^2 d\tau \over
\int \psi_0^2 d\tau}, \label{Heqn38}
\end{eqnarray}
we will be motivated to consider a DFT theory such as that
presented in section \ref{DFT} with appropriate approximations.
Note that the integral in eqn(\ref{Heqn38}) is over the
coordinates of {\it all} the electrons with the implicit
dependence of $R_{j,k}$ on the others. As can be easily seen, the
operator ${\cal L}_{j,k}$ simplifies considerably if the ground
state of the atom $\psi_g$ is well approximated by a Hartree type,
or KS type wavefunction for spherical atoms:
\begin{equation}
\psi_g = \prod_{j=1}^N \psi_j(r_j), \label{Heqn39}
\end{equation}
as in this case the eqns(\ref{Heqn35}) decouple. The accuracy of
the calculations will be dependent on approximations to the
density $f(r_j)$ and the wavefunction $\psi_0$. Further
investigations along these lines will be able to provide a
systematic study of van der Waals interactions as in the case of
hydrogen detailed here.
\section{Conclusion}
\label{Conclusion} This paper sets out an exact solution for the
van der Waals and other dispersion forces for the hydrogen
molecule using the method of Slater and Kirkwood,
\cite{SlaterKirkwood31}. By considering the density distributions
$f_{A,B,C}$ we have shown that in this case, an exact energy
density functional exists for this problem which when minimised
with respect to the density, yields the exact results. We have
shown that the energy constants $A,B,C$ can be calculated to any
desired accuracy, the first few decimals being in full agreement
with Pauling and Beach \cite{PaulingBeach35}. We have also
considered the extension of our method for more complex systems
such as hydrogenic systems and helium for which approximations
must be invoked.  A systematic study of these and generalizations
to include the effect of a surface \cite{ChoyByron95} will be the
subject of future work.

{\bf Acknowledgements:} The author wish to thank the NCTS (Taiwan)
for their hospitality during the final stages of this work. This
paper is dedicated to the memory of J. Mahanty, who crossed path
with the author during a period at the Australian National
University in the late 1980's.

\centerline{\bf Appendix}
{\bf Integral equation method}

The integral equation technique for solving the SK
eqn(\ref{Heqn8}) is due to Levi \cite{CourantHilbert}. We define
the operator $\cal W$ acting on any function such as $R$:
\begin{equation}
{\cal W}[R] \equiv \Delta R + a R_{\xi} + b R_{\xi^{\prime}} + c R
, \label{Heqn40}
\end{equation}
where the subscripts denote differentiation from hereon and
$\Delta$ is the two dimensional Laplacian operator, so that the SK
eqn(\ref{Heqn8}) is given by:
\begin{equation}
{\cal W}[R]= f ; \label{Heqn40a}
\end{equation}
in which:
\begin{eqnarray}
a = ({4\over \xi}-1) & , & \quad  b = ({4\over \xi^\prime}-1)
\nonumber
\\ c = -({1\over \xi}+{1\over \xi^\prime}) & , &\quad  f=1.
\label{Heqn41}
\end{eqnarray}
A uniqueness theorem for particular solutions can be proved for
any value of $c\leq 0$, \cite{CourantHilbert} thus guaranteeing a
solution for eqn(\ref{Heqn40a}). With the use of an appropriate
two dimensional Green's function $G(x,y|x^\prime,y^\prime)$ such
that:
\begin{equation}
\Delta G(x,y|x^\prime,y^\prime) = -2 \pi \delta(x-x^\prime)
\delta(y-y^\prime), \label{Heqn42}
\end{equation}
then it can be easily shown that for any arbitrary function
$\omega(x,y)$ the solution $R(x,y)$ is given by:
\begin{eqnarray}
R(x,y) &=& \omega(x,y)\nonumber \\ &+& \int dx^\prime \int
dy^\prime G(x,y|x^\prime,y^\prime) \rho(x^\prime,y^\prime).
\label{Heqn43}
\end{eqnarray}
The function $\rho(x,y)$ is given by the solution of the integral
equation:
\begin{equation}
\rho(x,y)=\int dx^\prime \int dy^\prime
K(x,y|x^\prime,y^\prime)\rho(x^\prime,y^\prime) + g(x,y) ,
\label{Heqn44}
\end{equation}
where the kernel $K(x,y|x^\prime,y^\prime)$ is of the form:
\begin{equation}
K(x,y|x^\prime,y^\prime) = {1\over 2\pi}\Bigl (a G_x + b G_y + c G
\Bigr ) , \label{Heqn45}
\end{equation}
and the function $g(x,y)$ is:
\begin{equation}
{1\over 2\pi}\Bigl ({\cal L}[\omega] - f \Bigr ). \label{Heqn46}
\end{equation}
That eqns(\ref{Heqn43}) to (\ref{Heqn46}) give a solution for
eqn(\ref{Heqn40a}) can be easily shown by operating on $R$ in
eqn(\ref{Heqn43}) with the operator ${\cal W}$. With an
appropriate choice of Green's function $G(x,y|x^\prime,y^\prime)$
and $\omega(x,y)$, the iteration of eqn(\ref{Heqn44}) is
equivalent to our solution for $R(x,y)$ as obtained in section
\ref{Exact}.


\begin{references}
\bibitem{comment1}Readers should note that there are two slightly
different definitions of Laguerre polynomials $L^m_n(x)$. We use
the one for physicists \cite{PaulingWilson35},\cite{Schiff68},
which differs from the one for mathematicians ${\tilde L}^m_n(x)$,
such as defined in Mathematica v 3.1 or in most table of
integrals. The difference is given by: $L^m_n(x)={\tilde
L}^m_{n-m}(x)$ which is a nasty source of errors as this author
has found.
\end{references}
\end{document}